\documentclass[twocolumn]{aastex61}
\usepackage{longtable}
\usepackage{float}

\usepackage{lineno}

\submitjournal{ApJ}

\shorttitle{Magnetic Fields in Planet-Hosting M-dwarf Stars}
\shortauthors{Wanderley et al.}
\begin{document}

\title{Magnetic Fields in a sample of planet-hosting M dwarf stars from Kepler, K2, and TESS observed by APOGEE}

\correspondingauthor{Fábio Wanderley}
\email{fabiowanderley@on.br}

\author[0000-0003-0697-2209]{Fábio Wanderley}
\affiliation{Observatório Nacional/MCTIC, R. Gen. José Cristino, 77, 20921-400, Rio de Janeiro, Brazil}

\author[0000-0001-6476-0576]{Katia Cunha}
\affiliation{Steward Observatory, University of Arizona, 933 North Cherry Avenue, Tucson, AZ 85721-0065, USA}
\affiliation{Observatório Nacional/MCTIC, R. Gen. José Cristino, 77,  20921-400, Rio de Janeiro, Brazil}

\author[0000-0002-0134-2024]{Verne V. Smith}
\affiliation{NSF’s NOIRLab, 950 N. Cherry Ave. Tucson, AZ 85719 USA}

\author[0000-0003-3061-4591]{Oleg Kochukhov}
\affiliation{Department of Physics and Astronomy, Uppsala University, Box 516, S-75120 Uppsala, Sweden}

\author[0000-0002-7883-5425]{Diogo Souto}
\affiliation{Departamento de F\'isica, Universidade Federal de Sergipe, Av. Marcelo Deda Chagas, S/N Cep 49.107-230, S\~ao Crist\'ov\~ao, SE, Brazil}

\author[0000-0002-0084-572X]{C. Allende Prieto}
\affiliation{Instituto de Astrofísica de Canarias, E-38205 La Laguna, Tenerife, Spain}
\affiliation{Departamento de Astrofísica, Universidad de La Laguna, E-38206 La Laguna, Tenerife, Spain}

\author{Suvrath Mahadevan}
\affiliation{Department of Astronomy \& Astrophysics, Pennsylvania State, 525 Davey Lab, University Park, PA 16802, USA}
\affiliation{Center for Exoplanets \& Habitable Worlds, Pennsylvania State, 525 Davey Lab, University Park, PA 16802, USA}

\author[0000-0003-2025-3147]{Steven R. Majewski}
\affiliation{Department of Astronomy, University of Virginia, Charlottesville, VA 22904-4325, USA}

\author[0000-0002-0638-8822]{Philip S. Muirhead}
\affiliation{Institute for Astrophysical Research and Department of Astronomy, Boston University, 725 Commonwealth Ave., Boston, MA 02215, USA}

\author[0000-0002-7549-7766]{Marc Pinsonneault}
\affiliation{Department of Astronomy, The Ohio State University, Columbus, OH 43210, USA}

\author{Ryan Terrien}
\affiliation{Department of Physics \& Astronomy, Carleton College, Northfield MN, 55057, USA}

\begin{abstract}
Stellar magnetic fields have a major impact on space weather around exoplanets orbiting low-mass stars. From an analysis of Zeeman-broadened Fe I lines measured in near-infrared SDSS/APOGEE spectra, mean magnetic fields are determined for a sample of 29 M dwarf stars that host closely orbiting small exoplanets. The calculations employed the radiative transfer code Synmast and MARCS stellar model atmospheres. The sample M dwarfs are found to have measurable mean magnetic fields ranging between $\sim$0.2 to $\sim$1.5 kG, falling in the unsaturated regime on the $<$B$>$ vs P$_{\rm rot}$ plane.
The sample systems contain 43 exoplanets, which include 23 from Kepler, nine from K2, and nine from TESS. We evaluated their equilibrium temperatures, insolation, and stellar habitable zones and found that only Kepler-186f and TOI-700d are inside the habitable zones of their stars. Using the derived values of $<$B$>$ for the stars Kepler-186 and TOI-700 we evaluated the minimum planetary magnetic field that would be necessary to shield the exoplanets Kepler-186f and TOI-700d from their host star's winds, considering reference magnetospheres with sizes equal to those of the present-day and young Earth, respectively.  Assuming a ratio of 5$\%$ between large-to-small scale B-fields, and a young-Earth magnetosphere, Kepler-186f and TOI-700d would need minimum planetary magnetic fields of, respectively, 0.05 and 0.24 G. These values are considerably smaller than Earth's magnetic field of 0.25 G$\lesssim$B$\lesssim$0.65 G, which suggests that these two exoplanets might have magnetic fields sufficiently strong to protect their atmospheres and surfaces from stellar magnetic fields.
\end{abstract}
\keywords{Near Infrared astronomy(1093) --- M dwarf stars(982) --- Stellar activity(1580) --- Stellar magnetic fields(1610)}

\section{Introduction}

M dwarfs are the most abundant stellar type in the Galaxy \citep{salpeter1955,reid1997} and also host the largest reservoir of small planets \citep{mulders2015,howard2012}.  Given their small sizes, M dwarfs are prime targets for the detection of Terrestrial-type planets, via both transit and radial-velocity surveys \citep{shields2016}. In addition, the low surface temperatures and luminosities of M dwarfs (L$_{bol}\lesssim$ 0.07 L$_{\odot}$) result in habitable zones (HZ), defined here as the region where liquid water can be sustained, that are much closer to the host star (e.g., \citealt{kopparapu2013}) when compared to the habitable zones of FGK dwarfs. However, the fact that a planet around an M dwarf is in the habitable zone does not mean that the planet is habitable \citep{gallet2017}.
An important facet of M-dwarf stars is that they have longer spin-down timescales than hotter main-sequence stars and they maintain intense magnetic fields for longer periods \citep{newton2016} that can, in turn, affect the evolution of planetary atmospheres and surfaces.

Stellar magnetic fields are a key ingredient in driving and shaping stellar winds, flare variability, and high-energy radiation (UV and X-ray) from cool dwarf stars with significant outer convective envelopes. The propagation of high-energy radiation and high-temperature plasma on the circumstellar environment away from the star, also referred to as space weather, impacts the environment of any orbiting planet \citep{shkolnik2014}. 
Several works have modeled stellar winds via magnetohydrodynamic (MHD) simulations to study space weather on M-dwarf planets (e.g., \citealt{vidotto2014,alvarado2019,harbach2021}). 
In particular, the relative strengths of the magnetic fields of the star and planet determine, for example, how important planetary atmospheric erosion will be, while changing magnetic field strengths will impact the planet's mass loss rates \citep{gupta2023}.

Ultimately, the magnetic field of the host star will have an impact on the possible habitability of planets. For example, previous works have modeled space weather around the M dwarf Proxima Centauri, which has a large-scale magnetic field (derived from Zeeman Doppler Imaging, based on circular polarimetry) $<$B$_{\rm ZDI}$$>$= 200 G \citep{klein2021}. 
\citet{garaffo2022}, using the magnetic-field determinations from \citep{klein2021}, found that the exoplanet Proxima Centauri b (semi-major axis of a=0.049 AU), which falls in the HZ, could support a planetary magnetosphere for planetary magnetic fields of B$_{\rm p}$=0.1 G, while \citet{alvarado2020} found an "Earth-like" stellar wind environment around the exoplanet Proxima Centauri c (a=1.44 AU), given the weak interplanetary magnetic field at the distance of this planet. However, at the distance of Proxima Centauri d (a=0.029 AU), the closest orbiting planet in the Proxima Cen system, the conditions were likely extremely adverse. Finally, habitability will also depend on planetary magnetic fields, as these can shield the atmosphere and the surface of the planet \citep{vidotto2013}.

In this study, we determine mean magnetic fields for a sample of 29 planet-hosting, partially-convective, M (and some late-K) dwarfs by modeling Zeeman intensified Fe I lines in SDSS APOGEE spectra \citep{majewski2017_apogee}. This stellar sample contains, in total, 23 exoplanets that were detected by the Kepler mission (\citealt{koch2010,borucki2010,batalha2013}), nine exoplanets detected by K2 \citep{howell2014}), nine exoplanets detected by TESS (Transiting Exoplanet Survey Satellite; \citealt{ricker2015}), one exoplanet detected with the HARPS spectrograph \citep{mayor2003} via the radial velocity method, and one exoplanet that was detected using data from multiple observatories via TTV (transit timing variation), for a total sample of 43 exoplanets. 

This paper is organized as follows: in Section 2 we present the sample of planet-hosting stars analyzed in this study. In Section 3 we discuss the methodology adopted for the mean magnetic field determination, and in Section 4, we present and discuss the results, which include an analysis of the derived magnetic fields as a function of rotational periods, and a discussion about the minimum planetary magnetic field necessary to deflect stellar coronal winds. 
Finally, Section 5 summarizes the conclusions. 

\section{Observations \& the Sample}

Our spectroscopic analysis is based on near-infrared ($\lambda$1.51$\mu$m to $\lambda$1.69$\mu$m), high-resolution (average resolution of R $\sim$ 22,500) spectra of M-dwarf stars observed by the SDSS III and IV APOGEE survey \citep{majewski2017_apogee,blanton2017}. The APOGEE spectra were obtained using the 300-fiber multi-object spectrographs located at two 2.5-m telescopes, with one at the Apache Point Observatory (APO) in the Northern Hemisphere, and the other at Las Campanas Observatory (LCO) in the Southern Hemisphere \citep{bowen1973,gunn2006_sdss,wilson2019}.

The stellar sample in this study was defined by cross-matching M dwarf stars from the APOGEE Data Release 17 (DR17) \citep{apogeedr17_2022} with T$_{\rm eff}$$<$4000 K, with the table of confirmed planets and host stars from the NASA Exoplanet Archive. We consider only stars 
unlikely to be members of binary or multiple systems by restricting the sample to those having values of Gaia DR3 \citep{gaia2022} RUWE $<$ 1.4 \citep{belokurov2020}. We also checked the literature to remove stars known to be members of binary or multiple systems, and, in addition, we examined the radial velocity scatter for stars that were observed more than once with APOGEE, removing stars with a radial velocity scatter greater than 1 km s$^{-1}$. 
The final stellar sample whose magnetic fields were determined in this study is composed of the 29 M dwarf stars presented in Table \ref{compiledata}. 
Within this final sample of 29 stars and 43 exoplanets, 22 of the stellar hosts have a single detected planet, while seven of them host more than one detected planet: Kepler-186 has five exoplanets, TOI-700 has four exoplanets, Kepler-138 has four exoplanets, and TOI-2095, Kepler-732, Kepler-1350 and K2-83 have two detected exoplanets each.

\section{Mean Magnetic Fields Determination}

The mean magnetic fields for the sample of M dwarfs were derived by employing the same methodology described in \citet{wanderley2024}. 
We measured the Zeeman effect in four selected Fe I  lines at $\lambda$15207.526 \r{A}, $\lambda$15294.56 \r{A}, $\lambda$15621.654 \r{A}, and $\lambda$15631.948 \r{A}, which have effective Landé-g factors of respectively 1.532, 1.590, 1.494 and 1.655, according to the VALD database \citep{piskunov1995,kupka1999}, making them sensitive lines to magnetic fields. The stellar parameters (effective temperatures, surface gravities, and stellar radii) for the sample were obtained using the procedure described in our previous studies \citep{souto2020,wanderley2023}. 
We derived projected stellar rotational velocities (v$\sin{i}$) using OH lines, keeping in mind that these are insensitive to magnetic fields (see \citealt{wanderley2024}).

Given the adopted stellar parameters, for each star we generated a spectral synthesis grid for the selected Fe I lines,  with $<$B$>$ values ranging from 2 to 6 kG with a 2 kG step, and metallicities [M/H] varying from -0.75 to +0.5 with a 0.25 step. This synthetic grid was constructed using the Synmast spectral synthesis code \citep{kochukhov2010_synmast}, the APOGEE DR17 line list \citep{smith2021}, and MARCS model atmospheres \citep{gustafsson2008_marcs}. Each synthesis was convolved with the derived stellar v$\sin{i}$ using a rotational profile, and with a Gaussian profile corresponding to the LSF of each fiber of the corresponding APOGEE spectrum (see, \citealt{wilson2019,nidever2015}).

We employed a Monte Carlo and Markov Chain (MCMC) methodology for the derivation of mean magnetic fields for the stars, using the python code emcee \citep{emcee2013} to search for the ensemble of filling-factors (varying $<$B$>$ between 2 -- 6 kG in a 2 kG step) and metallicity that best-fits the four Fe I lines. Each group of parameters has an associated $<$B$>$ value given by:

\begin{equation}
    <B>= \sum_{n} f_{n} \times 1000n \, \, , \, \, 
    n=[2,4,6] .
    \label{eqb}    
\end{equation}

\begin{figure*}
\begin{center}
  \includegraphics[angle=0,width=1\linewidth,clip]{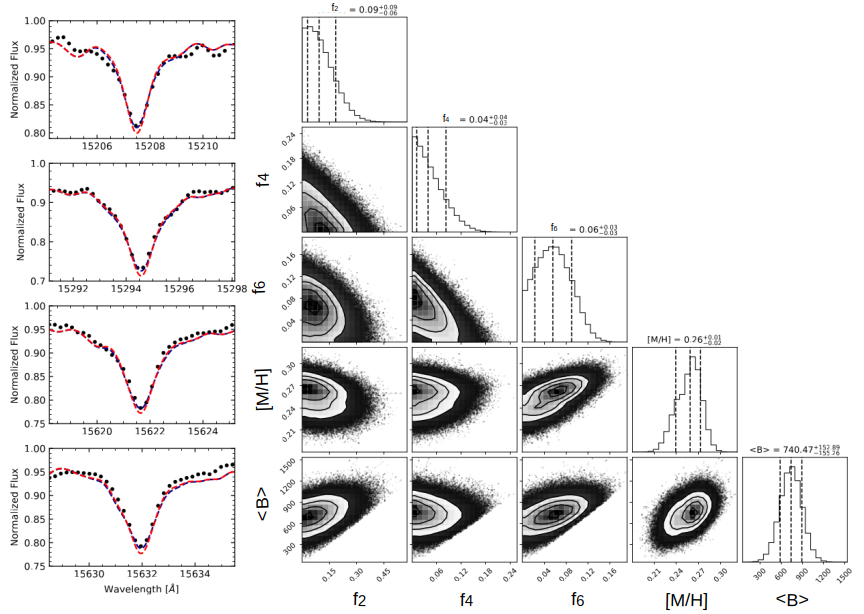}
\caption{Magnetic field analysis for the star 2M07590587+1523294. 
The left panels show the four Fe I lines used in the modeling: black dots are the observed APOGEE spectrum, red dashed lines are synthetic profiles computed without a magnetic field, and dark blue lines are our best fits obtained from the MCMC modeling. The right panel is a corner plot that presents the median and uncertainties (from $16^{th}$ and $84^{th}$ percentiles) of the derived parameters, which includes filling factors from magnetic fields between 2 and 6 kG in steps of 2 kG. We also show the final magnetic field result and its uncertainties.} 
\end{center}
\label{corner}
\end{figure*}

The mean magnetic fields were given by the median of the posterior distribution of $<$B$>$, with the lower and upper uncertainties given, respectively, by the $16^{th}$ and $84^{th}$ percentiles (Table \ref{compiledata}).
Figure \ref{corner} shows the results for the target M dwarf 2M07590587+1523294, where the left panels present comparisons between the model and observations, and best-fit results for the four studied Fe I lines are shown as blue lines. Black dots represent the observed APOGEE spectrum, and red lines are models computed with the exact same parameters as the blue lines, but without magnetic field. The right panel of Figure 1 shows the corresponding corner plot, with the posterior distribution for the filling factors and metallicity, as well as the posterior distribution for $<$B$>$. 

\section{Discussion}

\subsection{Stellar Magnetic Fields \& Stellar Rotation}

\begin{figure}
\begin{center}
  \includegraphics[angle=0,width=1\linewidth,clip]{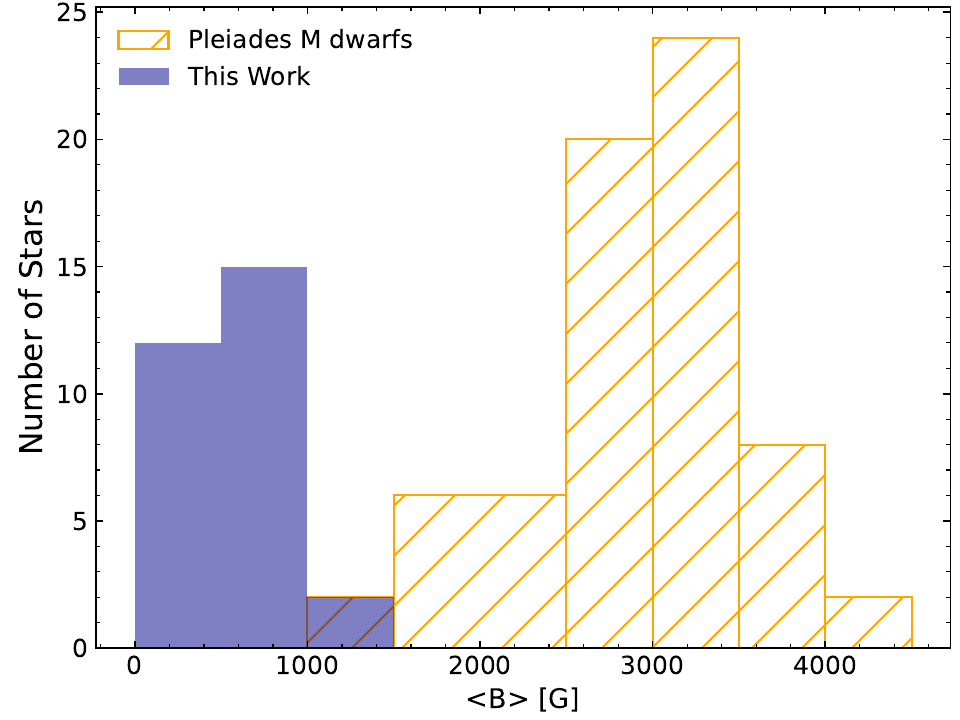}
\caption{The distribution of the derived mean magnetic fields for the sample of M dwarfs studied here (blue histogram), and for a sample of M dwarf stars belonging to the young Pleiades open cluster from \citet{wanderley2024} (orange histogram).}
\end{center}
\label{bhist}
\end{figure}

The mean magnetic fields for the studied M dwarfs range between $\sim$0.2 to $\sim$1.5 kG.
The distribution of the values for $<$B$>$ from this study is presented as the blue histogram in Figure \ref{bhist}. For comparison, we also show in this figure a histogram (in orange) representing the distribution of the mean magnetic fields obtained for the sample of M dwarf members of the young Pleiades open cluster studied by \citet{wanderley2024}, whose mean magnetic field results are on the same scale as those here. 
One can clearly see from the two distributions in Figure \ref{bhist} that there is an evolution in the mean magnetic fields between the two samples. 
For the Pleiades, where age$_{cluster}\sim$10$^{8}$yr, the magnetic fields for the M dwarfs are characterized by a distribution having stronger mean B-fields, with a peak at $\sim$ 3 kG. For the M dwarf sample of planet hosts in this study, on the other hand, the peak is at lower $<$B$>$ values, roughly between 0.5 and 1 kG, with only two stars having $<$B$>$ $>$ 1 kG.
The differences between the mean magnetic field distributions of the two samples as a function of the stellar rotational periods will be further discussed below. But we note here, that the planet-hosting field M-dwarfs in this study are likely much older than the Pleiades sample, based for example, on the different distributions of their rotational periods \citep{engle2023}.

\begin{figure}
\begin{center}
  \includegraphics[angle=0,width=1\linewidth,clip]{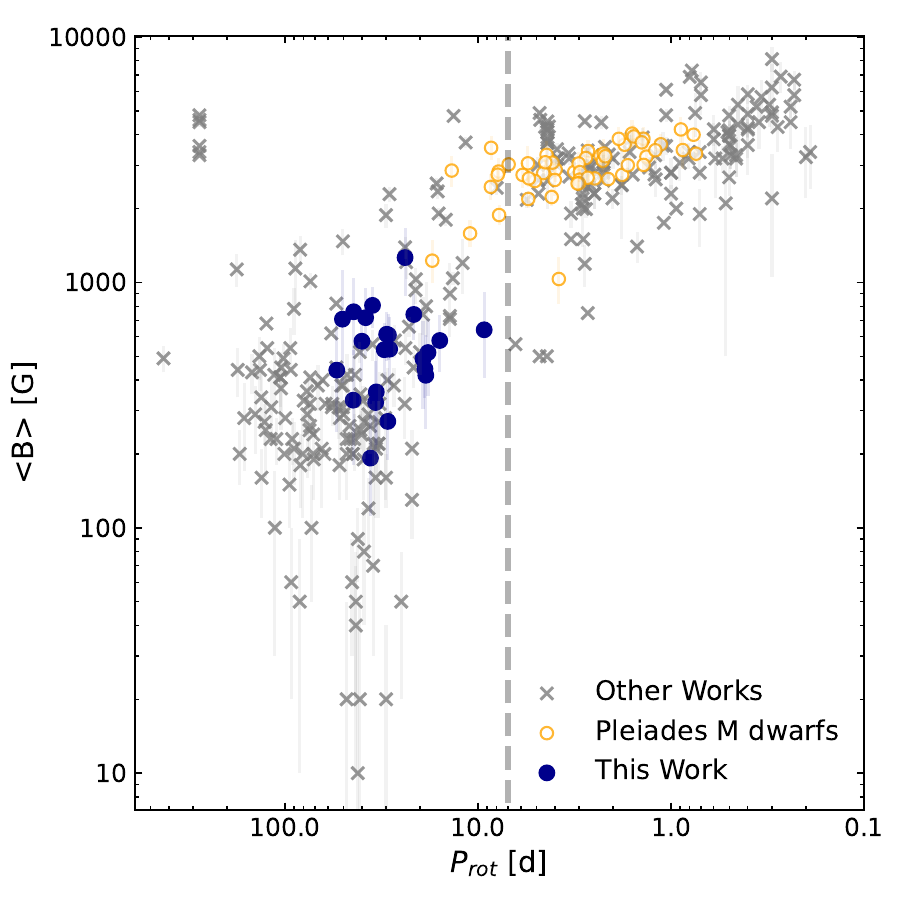}
\caption{Derived mean magnetic fields for the sample of planet-hosting M dwarfs as a function of rotational periods (filled blue circles). Other results for M dwarfs from the literature are also shown (grey xs, \citealt{saar1985_olegreview11, saar1994_olegreview12, saar1996_olegreview14, johns-krull1996_olegreview15, johns-krull2000_olegreview8, kochukhov2001_olegreview5, afram2009_olegreview9, phan-bao_olegreview13, kochukhov2009_olegreview6, shulyak2011_olegreview1, shulyak2014_olegreview10, kochukhov2017_olegreview4, shulyak2017_olegreview2, kochukhov2019_olegreview7, shulyak2019_olegreview3, reiners2022, cristofari2023a, cristofari2023b}), along with results for a sample of Pleiades M dwarf stars from \citet{wanderley2024} (2024; open orange circles). The grey dashed line at P$_{\rm rot}=7$ days represents the threshold that separates the saturated from the unsaturated regime. The mean magnetic fields for our sample fall in the unsaturated regime.}
\end{center}
\label{BPROT}
\end{figure}

The behavior of mean B-fields as a function of stellar rotational periods (on a log-log scale) is shown in Figure \ref{BPROT}. The literature magnetic field results exhibited in this figure (shown as grey xs) were all obtained from Zeeman split/intensified lines \citep{saar1985_olegreview11, saar1994_olegreview12, saar1996_olegreview14, johns-krull1996_olegreview15, johns-krull2000_olegreview8, kochukhov2001_olegreview5, afram2009_olegreview9, phan-bao_olegreview13, kochukhov2009_olegreview6, shulyak2011_olegreview1, shulyak2014_olegreview10, kochukhov2017_olegreview4, shulyak2017_olegreview2, kochukhov2019_olegreview7, shulyak2019_olegreview3, reiners2022, cristofari2023a, cristofari2023b}. The results from this study are shown as filled blue circles, with rotational periods taken from: \citet{mcquillan2013a,mcquillan2013b,mazeh2015,mann2017,torres2017,bonomo2017,kosiarek2019,reinhold2020,gilbert2020,bluhm2021,murgas2023,murgas2024}, and we also show, as open orange circles, mean magnetic field results for the Pleiades M dwarfs \citep{wanderley2024}.

Figure \ref{BPROT} exhibits an inverse correlation between rotational periods and mean magnetic fields, which is steeper for stars with P$_{\rm rot}>7$ days and flatter for stars with shorter rotational periods. 
The steeper and flatter relations between $<$B$>$ and P$_{rot}$ define two regimes; younger stars that rotate faster are in the 'saturated regime', where $<$B$>$ changes slowly as a function of P$_{\rm rot}$, and where the stars approach a limit of kinetic to magnetic energy conversion \citep{reiners2022, wanderley2023}, while older stars that rotate slower are in the 'unsaturated regime', where $<$B$>$ decreases with the rotational period.  Note that in the $<$B$>$ -- P$_{\rm rot}$ plane, the M dwarf sample studied here falls in the unsaturated regime; these stars have effective temperatures in the range between $\sim$3400 K to $\sim$4000 K and are partially convective. 
The grey dashed vertical line at P$_{\rm rot}=7$ days in Figure \ref{BPROT} represents the estimated threshold that separates the saturated and unsaturated regimes. We note again that, although we are using the same methodology to derive mean magnetic fields and analyzing M dwarfs within a similar effective temperature range as in the Pleiades study of \citet{wanderley2024}, the distribution of both results in the P$_{\rm rot}$ -- $<$B$>$ plane are quite different, with the field M dwarfs in this study being entirely in the unsaturated regime, while most Pleiades M dwarf stars are in the saturated regime.

\begin{figure}
\begin{center}
  \includegraphics[angle=0,width=1\linewidth,clip]{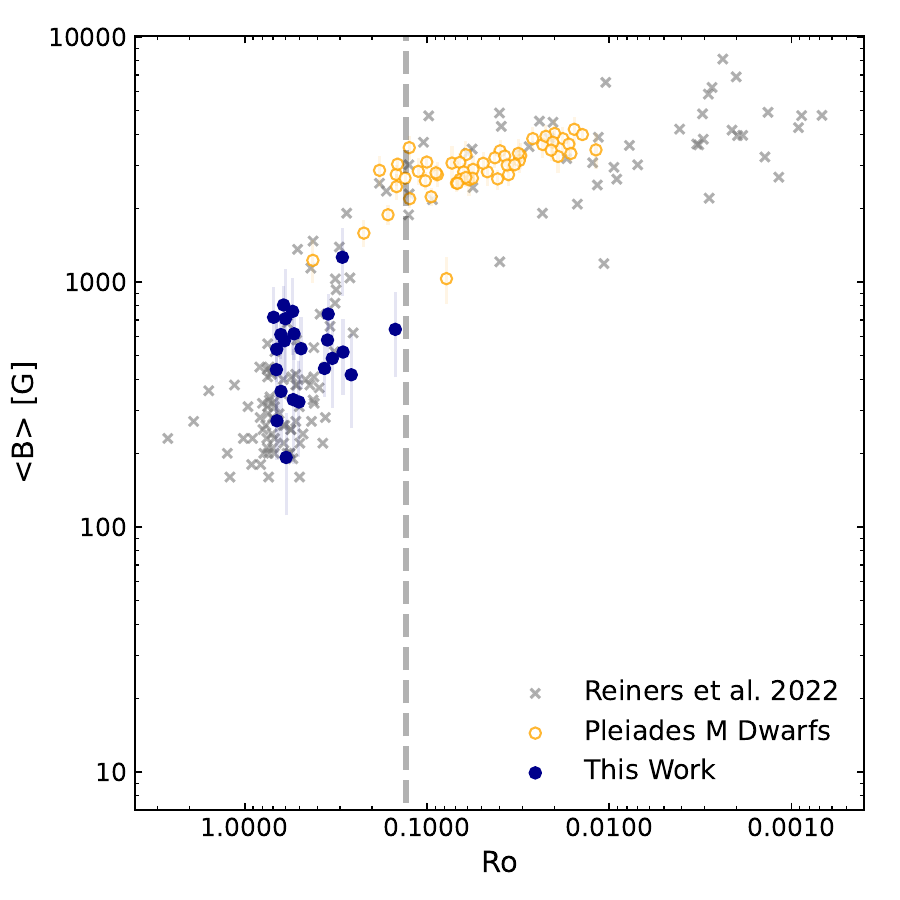}
\caption{Derived mean magnetic fields for the sample of planet-hosting M dwarfs as a function of Rossby numbers (filled blue circles). Results for M dwarfs from \citet{reiners2022} are shown (grey xs). The grey dashed line at Ro=0.13 represents the threshold that separates the saturated from the unsaturated regime.}
\end{center}
\label{bro}
\end{figure}

Figure \ref{bro} shows the relation between the derived mean magnetic fields as a function of Rossby numbers (filled blue circles). The Rossby number (Ro) is a dimensionless quantity that is given by the ratio between the stellar rotational period and convective turnover time ($\tau$, measured in days) and is an important activity indicator. We derived Rossby numbers adopting the same methodology employed by \citet{wanderley2024}, using the relation $\tau$ = 12.3 $\times$ (L/L$_{\odot}$)$^{-0.5}$ from \citet{reiners2022}. To derive luminosities we considered bolometric correction calibrations from \citet{mann2015,mann2016}, and distances from \citet{bailerjones2021}. In Figure \ref{bro} we show the results for field M dwarfs from \citet{reiners2022} (grey xs), with the grey dashed vertical line at Ro=0.13 in the figure, representing the estimated threshold that separates the saturated and unsaturated regimes. We see a similar behavior in the observed relation between mean magnetic fields and rotational periods and mean magnetic fields and Rossby numbers, indicating that the studied sample is composed of M dwarf stars with non-saturated magnetic fields, while the Pleiades M dwarf stars are mostly saturated.

The segregation between the Pleiades M dwarfs and the planet-hosting field M dwarfs can be explained by the age difference between the two samples. The Pleiades stars have ages around 100 Myr (age = 112 $\pm$ 5 Myr; \citealt{dahm2015_age4}), while the field star sample is likely much older and composed of stars that have dissipated much of their initial angular momentum, resulting in significantly smaller values of $<$B$>$.  \citet{engle2023} quantified age-rotational period relations for M dwarfs and, according to that study, the values of P$_{rot}$ $\sim$1-10 days that characterize the Pleiades sample span ages of one to a few $\times$10$^{8}$ yr, while the longer periods of P$_{rot}$ $\sim$9 -- 50 days that cover our planet-hosting field M-dwarf sample suggest ages of one to several Gyr.

\begin{figure}
\begin{center}
  \includegraphics[angle=0,width=1\linewidth,clip]{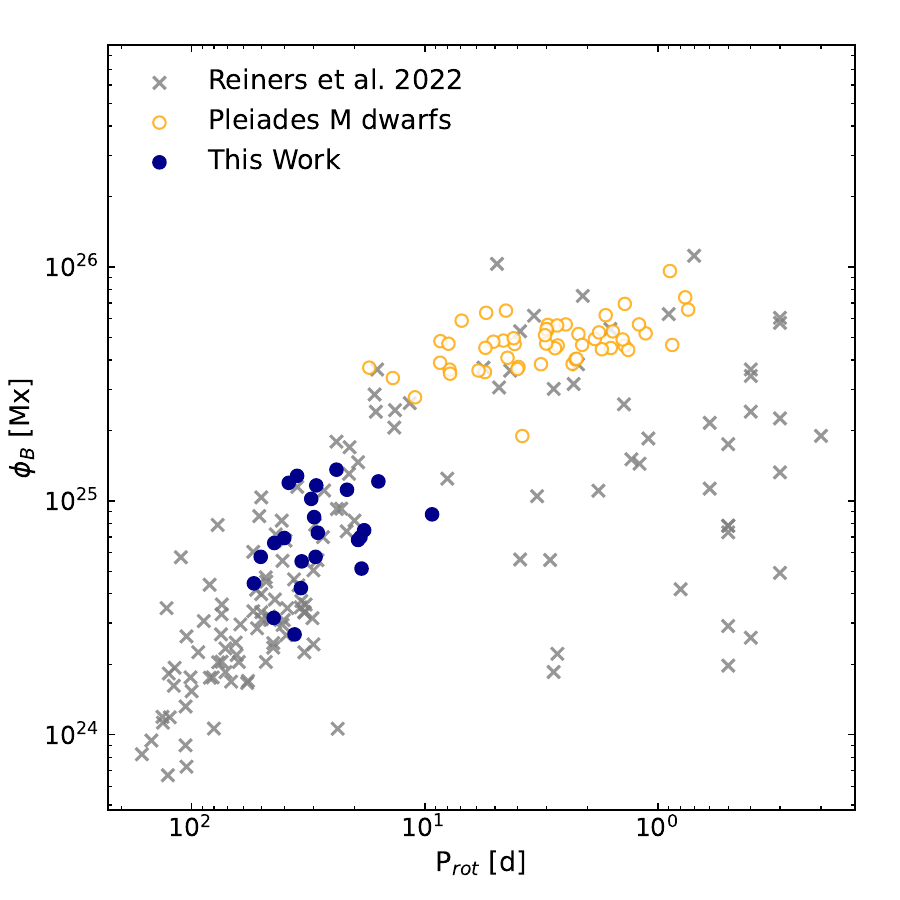}
\caption{Derived magnetic fluxes for the studied M dwarfs as a function of their rotational periods (blue circles). The M dwarf members of the young Pleiades open cluster (\citealt{wanderley2024}; open orange circles) and for field M dwarfs (\citealt{reiners2022}; grey xs) are also shown. The studied M dwarfs fall in the unsaturated regime and follow other results from the literature.}
\end{center}
\label{phib}
\end{figure}

We also derived magnetic fluxes in Mx units (1 Mx = 1 G $\times$ cm$^{2}$) for our sample by multiplying mean magnetic fields and the stellar area, given by $4 \pi R^{2}$. Figure \ref{phib} shows the magnetic fluxes for the sample stars as a function of their rotational periods as filled blue circles. 
We find an overall good agreement between the magnetic flux results in this study and those obtained for the field M dwarfs in \citet{reiners2022} (grey xs), within the overlapping range in rotational periods, while the magnetic fluxes for the young M dwarfs from the Pleiades open cluster have lower rotational periods and much higher magnetic fluxes. We also note that, as pointed out in \citet{wanderley2024}, the magnetic fluxes for the young Pleiades M dwarfs show much less scatter than those for the field stars from the literature, which, again, may be explained by the age difference between both samples, as the field stars are not necessarily expected to have similar ages.

\subsection{Planetary Habitability, Stellar \& Planetary Magnetic Fields}

As mentioned in the introduction, a star's habitable zone (HZ) is the region around it at which a planet would be able to maintain liquid water at its surface. Habitable zone calculations consider how planets process the thermal energy emitted by the host star. However, many other important features can drastically change habitability and ultimately significantly reduce the chances of finding life on a planet, in particular, for planets orbiting M dwarfs. Due to their lower thermal luminosity output, the habitable zones of exoplanets around M dwarfs are much closer to the star compared to those for hotter main-sequence stars. Therefore, there is a higher probability of tidal-locking in such systems.
Also, the typically high magnetic fields of M dwarf stars, which for some cases in this study can be $\sim$1 kG, are responsible for both Extreme-UV (EUV) emission and X-ray non-thermal emission \citep{wright2011,newton2017,astudillo2017,kochukhov2021,reiners2022,engle2024}
as well as intense stellar winds that may erode or even evaporate planetary atmospheres. A planetary magnetic field would then be needed for planetary protection \citep{zendejas2010,vidotto2013,see2014,nascimento2016,gallet2017,mozos2019,dimitri2021}. 

We start this discussion by investigating whether any of the exoplanets of the sample are inside the habitable zones of their stars. We used the 1D, radiative-convective, cloud-free climate model from \citet{kopparapu2013}. We adopted the 'runaway greenhouse limit' and 'maximum greenhouse limit' as, respectively, the inner (IHZ) and outer (OHZ) limits of the habitable zone. The model assumes H$_{2}$O- (IHZ) or CO$_{2}$- (OHZ) dominated atmospheres with N$_{2}$ as a background gas. In the runaway greenhouse limit, water cannot condense into liquid droplets, and the stratosphere is dominated by water vapor, which suffers photolysis from stellar X-ray and EUV radiation, causing the loss of hydrogen, which results in an efficient mechanism for water loss \citep{kasting1988}. 
In the maximum greenhouse limit, the atmosphere is dominated by CO$_{2}$, and the greenhouse effect is at its maximum, with each addition of CO$_{2}$, only increasing the albedo, and cooling the planet. To derive the greenhouse limits we used the equation below from \citet{kopparapu2014}:
\begin{equation}
    d = \bigg( \frac{L/L_{\odot}}{S_{\rm eff}} \bigg) ,
    \label{zhi}    
\end{equation}

\noindent where $d$ is the distance from the star given in AU for either IHZ or OHZ. S$_{\rm eff}$ is the effective flux which is given by the equation:

\begin{equation}
    S_{\rm eff} = S_{\rm eff \odot} + aT_{*} + bT_{*}^{2}+ cT_{*}^{3} + dT_{*}^{4} ,
    \label{seff}    
\end{equation}

\noindent where T$_{*}$ is the stellar effective temperature minus the solar effective temperature of 5780 K. The coefficients used in equation \ref{seff}, as well as S$_{\rm eff \odot}$, are given in Table 1 of \citet{kopparapu2014}, with the runaway greenhouse limits depending on the planetary mass, and coefficients given for three planetary masses of 0.1M$_{\oplus}$,  1M$_{\oplus}$, and 5M$_{\oplus}$. 
(As discussed by \citet{kopparapu2014}, the habitable zone model depends on features such as planetary mass and the amount of background N$_{2}$ gas, and further work with 3D climate models is needed to improve the calculation for habitable zones around different types of stars).

We computed habitable zones for those exoplanets in our sample with available semi-major axis measurements ($a$) in the literature and found that only two of them lie inside their habitable zones. One of these exoplanets is the well-known planet Kepler-186f \citep{quintana2014}, which is located at 0.386 AU from its host star and sits comfortably inside its host star's habitable zone, with OHZ = 0.452 AU and IHZ = 0.229 -- 0.252 AU; we derived for this exoplanet a radius of R$_{\rm p}$=1.18 R$_{\oplus}$ (the same value from the DR25 KOI Table, \citealt{thompson2018}), 
and depending on its planetary mass, Kepler-186f may be similar to the Earth. The other exoplanet from our sample that is in the habitable zone is TOI-700d \citep{gilbert2020}. For this exoplanet, we derived a radius of R$_{\rm p}$= 1.16 R$_{\oplus}$ (similar to the value of R$_{\rm p}=1.073$ R$_{\oplus}$ from \citealt{gilbert2023}), and it has a distance from its host star of 0.163 AU. If we consider that TOI-700d has a mass between 1 -- 5 M$_{\oplus}$, it would be inside the habitable zone with IHZ = 0.153 -- 0.158 AU. If, on the other hand, this planet were much less massive, it would be out of the habitable zone, given that, for M$_{\rm p}$= 0.1M$_{\oplus}$, IHZ = 0.168 AU. 

We can also compute the insolation, S$_{\rm p}$, i.e., the energy flux that a planet receives from its host star, using the equation

\begin{equation}
    \frac{S_{\rm p}}{S_{\oplus}}  = \bigg( \frac{R_{*}}{R_{\odot}} \bigg)^{2} \bigg( \frac{T_{\rm eff *}}{T_{\rm eff \odot}} \bigg)^{4} \bigg( \frac{1AU}{a} \bigg)^{2} ,
    \label{sp}    
\end{equation}

\noindent where R$_{*}$ is the stellar radius. 

Another important exoplanet parameter is the equilibrium temperature T$_{\rm eq}$, which is the temperature of the planet considering that all the absorbed energy is re-emitted as thermal radiation, i.e., the planet is neither cooling nor heating. The equilibrium temperature only considers heating by stellar radiation, and it does not account for other heating or cooling mechanisms such as, for example, the greenhouse effect. 
The planet's equilibrium temperature is given by

\begin{equation}
    T_{\rm eq}  = T_{\rm eff}(1-A)^{0.25} \sqrt{\frac{R_{*}}{2a}} ,
    \label{teq}    
\end{equation}

\noindent where A is the planetary albedo, which we adopt here to be the Earth albedo of A$_{\oplus}$=0.3 \citep{goode2001}.
The derived habitable zones, insolation levels, and equilibrium temperatures, along with the orbital semi-major axis of the studied exoplanets are presented in Table \ref{compiledata}.

\begin{figure}
\begin{center}
  \includegraphics[angle=0,width=1\linewidth,clip]{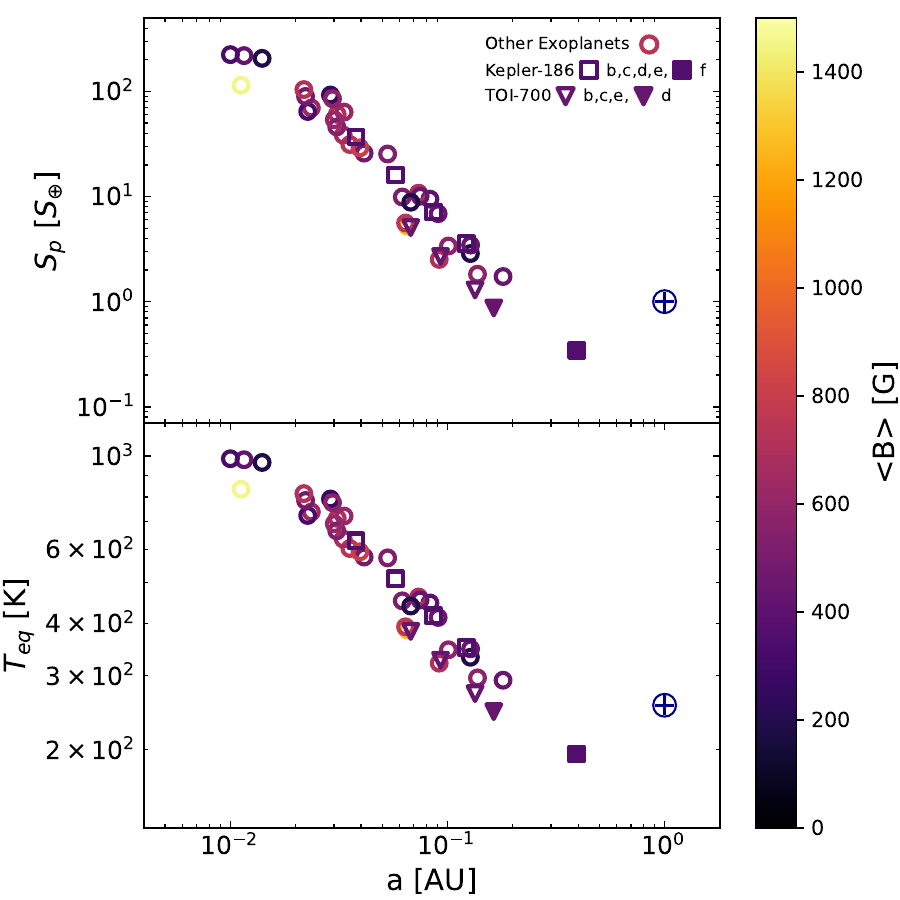}
\caption{The top and bottom panels show, respectively, the insolation and equilibrium temperatures as a function of the orbit semi-major axis for the sample exoplanets. The exoplanets are color-coded by the host star mean magnetic fields derived in this work; these are used to calculate the minimum planetary mass needed to protect the exoplanet from stellar winds. Most of the sample exoplanets are not in a habitable zone and these are shown as open symbols. The only exoplanets in the habitable zone of their stars are Kepler-186f (filled square) and TOI-700d (filled inverted triangle); the Earth is also shown for comparison.
}
\end{center}
\label{habit}
\end{figure}

In Figure \ref{habit} we combine the quantities derived above for the studied exoplanets and show the insolation (top panel) and equilibrium temperatures (bottom panel) as a function of the semi-major axis of the orbit of the exoplanet. The exoplanets are color-coded according to the mean magnetic field of the host stars in each case (see color bar). 
Most of the planets in our sample are closer to their host stars than the IHZ, which is probably related to detection biases since both the radial velocity and transit methods are much more sensitive to detecting planets that are closer to their host stars. The only two planets in our sample found to be in the habitable zone, Kepler-186f and TOI-700d (shown as filled symbols), are the ones suffering the lowest insolation and having the lowest equilibrium temperatures. In contrast, exoplanets in our sample that are closer to their stars receive higher insolation and have higher equilibrium temperatures, which contribute to them being out of the HZ. 

Our measurements of the mean magnetic fields for the two stars in this sample that host planets in the HZ, Kepler-186, and TOI-700, can be used to estimate the impact of a magnetically driven wind on both Kepler-186f and TOI-700d.  \citet{vidotto2013} investigated the effect of magnetic fields from M dwarf stars on the space environments around their planets and derived equations to estimate the minimum planetary magnetic fields required to deflect stellar wind pressure that is driven by stellar magnetic fields. We adopt the precepts of \citet{vidotto2013} to estimate the minimum planetary B-fields that would be needed to shield their surfaces, especially their atmospheres, from M dwarf winds. We begin with the expression derived by \citet{vidotto2013} for P$_{\rm B*}$(a), the magnetic pressure as a function of the orbital semi-major axis $a$
(their equation 6):

\begin{equation}
    P_{\rm B*}(a)= \frac{B_{\rm SS}^{2}}{8 \pi} \bigg( \frac{R_{\rm SS}}{a} \bigg)^{4} ,
    \label{vid6}    
\end{equation}

\noindent where B$_{\rm SS}$ is the stellar magnetic field evaluated at the source surface radius, R$_{SS}$. \citet{vidotto2013} derived this expression using the potential field source surface (PFSS) method, which is valid for a$>$R$_{\rm SS}$, and assumes that the stellar magnetic field decays with the square of the distance.

For a planetary surface or atmosphere to be shielded from a magnetized stellar wind, a sufficiently strong planetary magnetic field will be needed to balance the magnetic pressure from the stellar wind at some point away from the planet, which we call the magnetospheric radius, r$_{\rm M}$. In the case of rocky terrestrial-type planets that orbit M dwarfs, it is expected that stellar magnetic pressure will dominate over stellar-wind ram pressure, which will be ignored (e.g., \citealt{vidotto2011}). The equation below then describes the planetary magnetic pressure at a distance r$_{\rm M}$ from the planetary center (equation 1 from \citealt{vidotto2013}):

\begin{equation}
    P_{\rm Bp}(r_{\rm M})= \frac{[B_{\rm p}(r_{\rm M})]^{2}}{8 \pi} .
    \label{pbprm}    
\end{equation}

We assume a dipole field for a planetary magnetic field so that the field strength, B$_{\rm p}$, at a distance r$_{M}$ from the planetary center will have the form

\begin{equation}
    B_{\rm p}(r_{\rm M})= \frac{B_{\rm p0}}{2} \bigg( \frac{R_{\rm p}}{r_{\rm M}} \bigg)^{3} ,
    \label{bprm}    
\end{equation}

\noindent where R$_{\rm p}$ is the planetary radius and B$_{\rm p0}$ is the planetary magnetic field strength at the pole. For the planetary magnetic field pressure to balance the stellar wind magnetic pressure at a distance, $a$, from the star and a distance, r$_{\rm M}$, from the planet, then P$_{\rm Bp}$($r_{\rm M}$)= P$_{\rm B*}$(a) and an expression for the requisite planetary magnetic field is:       

\begin{equation}
    B_{\rm p,min}= 2B_{\rm SS} \bigg( \frac{R_{\rm SS}}{a} \bigg)^{2} \bigg( \frac{r_{\rm M}}{R_{\rm p}} \bigg)^{3} .
    \label{bp0}    
\end{equation}

In equation \ref{bp0}, R$_{\rm SS}$, the launching radius for the stellar wind, is set to 2.5R$_{*}$, as discussed in \citet{vidotto2013}, and the stellar magnetic field must be evaluated at R$_{\rm SS}$. The mean stellar magnetic fields presented here are derived from Zeeman-broadened Fe I lines, which map small-scale fields integrated over the stellar surface, while the magnetic fields at 2.5R$_{*}$ are expected to be dominated by the large-scale stellar magnetic field. As discussed by \citet{vidotto2011} at the position of a planet, the large-scale component of the magnetic field (obtained from ZDI, for example) survives, while the small-scale magnetic fields (our determinations) do not. We estimated $<$B$_{\rm ZDI}$$>$ from our derived $<$B$>$ for the star, considering the fraction $f$=$<$B$_{\rm ZDI}$$>$/$<$B$>$ of these two stellar magnetic field characteristics derived in previous studies. \citet{kochukhov2021} provides a detailed review of magnetic fields in M dwarf stars and compiles a list of stars with both small- and large-scale magnetic field measurements. As can be seen in Figure 15 in their work, for stars with $<$B$>$ and mass values in the range of our sample ($<$B$>$ $\lesssim$ 1.5 kG, and $M_{*}$ $\gtrsim$ 0.4 M$_{\oplus}$), the ratio between large- and small-scale magnetic fields never reaches $\sim$10$\%$. 
Here, we examine a range of possibilities by determining B$_{\rm p,min}$ using f=1$\%$, f=5$\%$ and f=10$\%$.  We also assume a dipole-like behavior of the value of B$_{\rm SS}$ at 2.5R$_{*}$, so that B measured at R=2.5R$_{*}$ will be 0.051$\times$ $<$B$>$ at the stellar surface. Inserting these factors into equation \ref{bp0} results in the expression:

\begin{equation}
    B_{\rm p,min}= 0.102 f \times <B> \bigg( \frac{2.5R_{*}}{a} \bigg)^{2} \bigg( \frac{r_{M}}{R_{\rm p}} \bigg)^{3} .
    \label{bp02}    
\end{equation}

This expression can now be evaluated for the different values of $f$ discussed above, as well as for different planetary magnetospheric radii, r$_{\rm M}$/R$_{\rm p}$.  We considered magnetospheric radii based on the size of the Earth's magnetosphere today and 3.4 Gyr ago, which are respectively r$_{\rm M}$/R$_{\rm p}$=11.7 and r$_{\rm M}$/R$_{\rm p}$=5 \citep{tarduno2010}.
The derived B$_{\rm p,min}$ for all sample exoplanets are presented in Table \ref{compiledata}.

\begin{figure}
\begin{center}
  \includegraphics[angle=0,width=1\linewidth,clip]{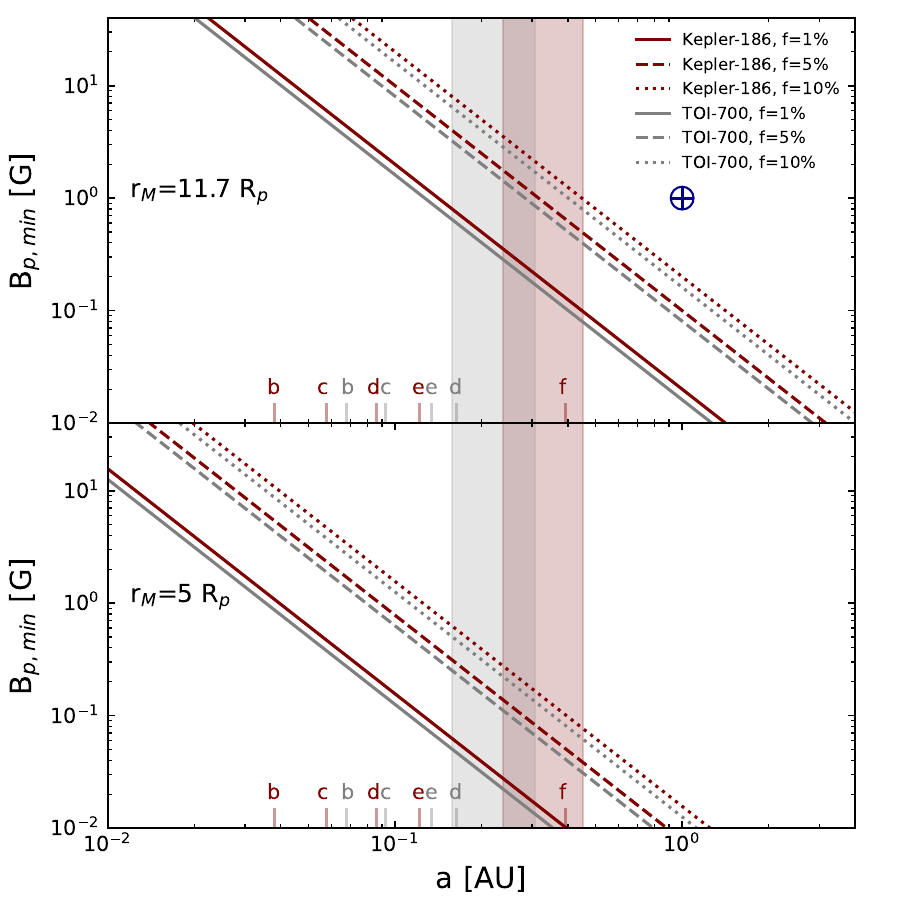}
\caption{Minimum planetary magnetic fields for stellar winds deflection (B$_{\rm p,min}$) as a function of orbit semi-major axis ($a$). The top and bottom panels consider, respectively, magnetospheric sizes of r$_{\rm M}$=11.7 R$_{\rm p}$ (Earth today) and r$_{\rm M}$=5 R$_{\rm p}$ (Earth, 3.4 Gyr ago). The maroon and grey colors are associated with planetary systems of the stars Kepler-186 and TOI-700, respectively. The small vertical lines at the bottom of each panel show the semi-major axis of the exoplanets around these stars. The shaded regions represent the habitable zones for each star, with IHZ determinations based on a planetary mass of 1 M$_{\oplus}$. Solid, dashed, and dotted lines are the B$_{\rm p,min}$ versus $a$ for each star, considering ratios between magnetic fields from circular polarimetry and intensity of respectively, 1$\%$, 5$\%$, and 10$\%$. We also show the position of the Earth in the upper panel for comparison.} 
\end{center}
\label{pblfig}
\end{figure}

Figure \ref{pblfig} shows the minimum magnetic planetary field, B$_{\rm p,min}$, as a function of the exoplanet semi-major axis, considering a present-day (top panel) and a young (bottom panel) Earth-size magnetosphere, for the two stars in our sample that host exoplanets inside habitable zones, Kepler-186 (represented in maroon) and TOI-700 (represented in grey). The small vertical lines and corresponding designating letters show the positions of the planetary systems in terms of the semi-major axis of their orbits. The maroon and grey shaded regions in the figure are the habitable zones of  Kepler-186 and TOI-700, respectively, considering a IHZ for 1M$_{\oplus}$ planetary mass. 
The inclined solid, dashed, and dotted lines are the B$_{\rm p,min}$ versus $a$ relations computed using three different values for f (the ratio between magnetic fields obtained from ZDI to Zeeman enhanced spectral lines) of $f$=1$\%$, $f$=5$\%$, and $f$=10$\%$. As previously concluded, we can see that only the exoplanets Kepler-186f and TOI-700d are inside the habitable zones of the stars, with the other exoplanets being too hot to sustain liquid water on the surface. For comparison, we show the position of the Earth in the upper diagram.

For a present-day Earth magnetosphere, assuming r$_{\rm M}$/R$_{\rm p}$=11.7 (top panel of Figure \ref{pblfig}), and taking the results for Kepler-186 of $<B>$ = 358 G, along with R$_{*}$ = 0.50 R$_{\odot}$, and the distance for exoplanet Kepler-186f of a=0.386 AU, we find B$_{\rm p,min}$ values for f=1$\%$, f=5$\%$ and f=10$\%$ of, respectively, 0.13 G, 0.65 G and 1.29 G, indicating that for all cases, Kepler-186f would in principle be able to sustain an earth-sized magnetosphere if it had a magnetic field similar to the Earth.
Doing the same estimate for the exoplanet TOI-700d, where $<$B$>$=439 G, R$_{*}$ = 0.41 R$_{\odot}$, and a = 0.163 AU, we obtain for f=1$\%$, f=5$\%$, and f=10$\%$, B$_{\rm p,min}$ values of respectively 0.60 G, 3.02 G, and 6.04 G. 
TOI-700d would require larger minimum planetary magnetic fields than Kepler-186f and, in particular for the f=5$\%$, and f=10$\%$ cases, B$_{\rm p,min}$ is of a few times the magnetic field of the Earth.

If we now take the magnetosphere of the young Earth, around 3.4 Gyr ago, with a magnetospheric size of r$_{\rm M}$/R$_{\rm p}$=5 (bottom panel of Figure \ref{pblfig}), we obtain for Kepler-186, B$_{\rm p,min}$ values for f=1$\%$, f=5$\%$ and f=10$\%$, of respectively 0.01 G, 0.05 G and 0.1 G.
For these f ratios, it is not unreasonable to assume that the magnetic field of the exoplanet Kepler-186f is probably able to protect the planet and deflect coronal material from the host star. For TOI-700d we obtain a similar conclusion, with B$_{\rm p,min}$ values for f=1$\%$, f=5$\%$, and f=10$\%$, of respectively 0.05 G, 0.24 G, and 0.47 G.

As a final note, it is important to keep in mind that these estimations are based on a steady state, large-scale B-field that drives stellar winds; stellar flares, large coronal mass ejections, and intense non-thermal radiation such as EUV and X-rays, which will impact habitability and planetary atmospheric and surface evolution, have not been considered. 

\section{Conclusions}

As M-dwarf stars are the most abundant stellar type in our Galaxy and known to host more terrestrial-sized planets than their warmer main-sequence stellar siblings, they are interesting candidates in the search for planets having conditions that could possibly support life. Due to observations of strong magnetic fields around many M dwarfs, coupled with habitable zones (i.e., the region around a host star where liquid water can be sustained) that lie in close proximity to their host stars, stellar magnetic characterization of M dwarfs is a crucial step for understanding possible magnetic interactions between host stars and their planets.  Magnetic fields are responsible for generating intense non-thermal EUV and X-ray radiation, as well as driving and controlling stellar winds, two processes that can impact the atmospheric evolution of orbiting planets \citep{owen2012, galian2023}. 
 
In our previous study of M-dwarf members of the young Pleiades open cluster \citep{wanderley2024}, we showed that the high-resolution (R=22,500), near-infrared (1.51 -- 1.69 micron) SDSS APOGEE spectra could be used, in conjunction with the radiative transfer code Synmast \citep{kochukhov2010_synmast}, to derive mean B-field strengths in these cool stars via Zeeman broadening of magnetically-sensitive Fe I lines. Here, we extend this type of analysis to a sample of 29 planet-hosting M-dwarf field stars observed by the APOGEE survey, all of which have effective temperatures between $\sim$3400 K and $\sim$4000 K. 
This particular M-dwarf sample hosts a total of 43 exoplanets, with 23 exoplanets detected by Kepler, nine exoplanets by K2, nine exoplanets by TESS, one exoplanet detected with the HARPS spectrograph, and another via TTV. 

The results of our analysis of Zeeman-broadened lines found that our sample of M-dwarf planet hosts has measurable magnetic fields, with mean $<$B$>$-fields ranging between $<$B$>$$\sim$0.2 to 1.5 kG. These mean magnetic fields for the sample stars are significantly weaker when compared to the ones obtained for the Pleiades M dwarfs in \citep{wanderley2024}. 
An investigation into the relation between mean magnetic fields, $<$B$>$, as a function of stellar rotational periods (P$_{\rm rot}$) found that all of the stars from the planet-hosting field-star sample with P$_{\rm rot}$ measurements lie in the non-saturated regime. This is in contrast to the results for the Pleiades M dwarfs, where a large majority falls within the saturated area, and exhibits stronger mean B-fields of $<$B$>$$\sim$ 2-- 4 kG, which can be explained by the young age of the cluster (t$_{cluster}\sim$10$^{8}$yr). 
The location of the Pleiades M-dwarfs and the planet-hosting field M-dwarfs studied here all fall within a locus in $<$B$>$ versus P$_{\rm rot}$ defined by B-field measurements for other M dwarfs taken from the literature, suggesting that the planet-hosting M dwarfs do not have atypical mean magnetic fields. 
Similarly to mean magnetic fields, computed values for the mean magnetic fluxes, $<$$\phi_{B}$$>$, for the studied M dwarfs also correlate with P$_{\rm rot}$, and also define both unsaturated and saturated regimes, in agreement with published values from the literature. 
 
We also analyzed habitability for our sample of exoplanets by calculating the locations of their habitable zones (using prescriptions from \citealt{kopparapu2013}), as well as equilibrium temperatures and insolation levels. 
We find that only two of the exoplanets in our sample are inside the habitable zones of their respective stars, Kepler-186f and TOI-700d. 
The remaining planets are outside, in regions with greater insolation levels and equilibrium temperatures that are too hot for water to condense into a liquid. However, although Kepler-186f and TOI-700d are inside their respective HZs, 
there are other effects aside from stellar thermal radiation that can compromise habitability in these systems, such as the effect of magnetic fields from the host M dwarfs.

Stellar magnetic fields provide important constraints on planet habitability. The mean stellar magnetic fields determined in this study were used to derive the minimum planetary magnetic field needed to protect the exoplanets from magnetically-controlled stellar winds coming from the host M-dwarf star.
We considered two scenarios based on \citet{vidotto2013}'s framework, one that uses as a reference a present-day Earth magnetospheric size of 11.7 R$_{\rm p}$, and another that considers a magnetosphere of 5 R$_{\rm p}$, corresponding to a young Earth some $\sim$3.4 Gyr ago.

In particular, we evaluated minimum planetary magnetic fields, B$_{\rm p,min}$, for those exoplanets in our sample that were inside their habitable zones, Kepler186-f and KOI-700d.  We investigated B$_{\rm p,min}$ assuming three different ratios ($f$ = 1$\%$, 5$\%$, and 10$\%$), between the large-scale magnetic field derived from circular polarimetry, $<$B$_{\rm ZDI}$$>$, which survives at the exoplanet's position, and the small-scale magnetic field that we derive from Zeeman broadened lines, $<$B$>$, which does not survive at large distances from the stellar surface. 

Assuming a magnetosphere corresponding to the present-day Earth, and a reasonable value for $f$ of 5$\%$ \citep{morin2010,vidotto2013,kochukhov2021}, Kepler-186f would need a planetary magnetic field of 0.65 G to resist atmospheric stripping from stellar winds, which is smaller than the Earth's present-day magnetic field of 0.25 G$\lesssim$B$\lesssim$0.65 G \citep{finlay2010}. The situation for the exoplanet TOI-700d, on the other hand, would be more difficult, as it would need a
magnetic field of 3.02 G, which is around 3 times the Earth's magnetic field, to protect the planet. If, however, we adopt the early Earth magnetosphere r$_{\rm M}$=5 R$_{\rm p}$ as a reference, these values change for Kepler-168f and TOI-700d to, respectively, 0.05 G and 0.24 G, meaning that in this scenario both exoplanets would probably be able to shield their atmosphere from stellar winds. 
As a point of interest, 3.4 Gyr ago on the Earth there was an already established complex biological carbon cycle in a marine ecosystem existing within this smaller magnetosphere \citep{reinhardt2024}, which suggests that life might also develop in exoplanets in the habitable zone having such a magnetospheric size.

In conclusion, the results of this study suggest that the exoplanets Kepler-186f and TOI-700d possibly have the conditions necessary to maintain liquid water on their surfaces. In addition, given the measured mean magnetic fields of their host M dwarfs, the required planetary magnetic field needed to be able to shield these exoplanets from their host star winds due to intense magnetic fields is modest and not unreasonable for Earth-like planets, potentially making Kepler-186f and TOI-700d relevant for future exoplanetary habitability studies.

\startlongtable
\begin{deluxetable*}{lcccccccccccc}
\small\addtolength{\tabcolsep}{-2pt}
\tablenum{1}
\label{compiledata}
\tabletypesize{\scriptsize}
\tablecaption{Stellar and Planetary Results }
\tablewidth{0pt}
\tablehead{
\colhead{APOGEE ID} &
\colhead{Exoplanet} &
\colhead{$<$B$>$} &
\colhead{$\phi_{\rm B}$} &
\colhead{P$_{\rm rot}$} &
\colhead{OHZ} &
\colhead{IHZ$^{a}$} &
\colhead{a} &
\colhead{T$_{\rm eq}$} &
\colhead{S$_{\rm p}$} &
\colhead{B$_{\rm p,min}^{b}$} &
\colhead{B$_{\rm p,min}^{c}$} & \\
\colhead{...} &
\colhead{...} &
\colhead{G} &
\colhead{$10^{25}$ Mx} &
\colhead{day} &
\colhead{AU} &
\colhead{AU} &
\colhead{AU} &
\colhead{K} &
\colhead{S$_{\oplus}$} &
\colhead{G} &
\colhead{G} &
}
\startdata
2M00391724+0716375 & K2-149 b & 402$_{-115}^{+122}$ & 0.813 & … & 0.503 & 0.264 & 0.083 & 447 & 9.45 & 0.33/1.67/3.34 & 4.28/21.41/42.83 \\
2M03593637+1533320 & K2-83 b & 695$_{-217}^{+230}$ & 1.119 & … & 0.473 & 0.249 & 0.031 & 711 & 60.51 & 3.28/16.42/32.85 & 42.09/210.43/420.86 \\
2M03593637+1533320 & K2-83 c & 695$_{-217}^{+230}$ & 1.119 & … & 0.473 & 0.249 & 0.074 & 462 & 10.77 & 0.58/2.92/5.84 & 7.49/37.44/74.89 \\
2M04342248+4302148 & TOI-1685 b & 418$_{-165}^{+194}$ & 0.513 & 18.66 & 0.344 & 0.178 & 0.012 & 981 & 219.23 & 10.88/54.41/108.81 & 139.42/697.12/1394.24 \\
2M04545692-6231205 & TOI-206 b & 1458$_{-410}^{+402}$ & 1.017 & … & 0.243 & 0.125 & 0.011 & 834 & 114.62 & 22.97/114.85/229.71 & 294.32/1471.6/2943.2 \\
2M06282325-6534456 & TOI-700 d & 439$_{-194}^{+248}$ & 0.443 & 54 & 0.307 & 0.158 & 0.163 & 246 & 0.87 & 0.05/0.24/0.47 & 0.6/3.02/6.04 \\
2M06282325-6534456 & TOI-700 c & 439$_{-194}^{+248}$ & 0.443 & 54 & 0.307 & 0.158 & 0.093 & 326 & 2.69 & 0.15/0.73/1.46 & 1.87/9.33/18.65 \\
2M06282325-6534456 & TOI-700 b & 439$_{-194}^{+248}$ & 0.443 & 54 & 0.307 & 0.158 & 0.068 & 382 & 5.06 & 0.27/1.37/2.74 & 3.51/17.56/35.12 \\
2M06282325-6534456 & TOI-700 e & 439$_{-194}^{+248}$ & 0.443 & 54 & 0.307 & 0.158 & 0.134 & 272 & 1.29 & 0.07/0.35/0.7 & 0.9/4.48/8.96 \\
2M07590587+1523294 & GJ 3470 b & 740$_{-155}^{+153}$ & 1.115 & 21.54 & 0.397 & 0.206 & 0.036 & 603 & 31.23 & 2.51/12.53/25.06 & 32.11/160.54/321.07 \\
2M08255432+2021344 & K2-122 b & 271$_{-83}^{+86}$ & 0.576 & 29.37 & 0.544 & 0.287 & 0.029 & 791 & 92.53 & 1.95/9.76/19.53 & 25.02/125.09/250.18 \\
2M08372705+1858360 & K2-95 b & 1262$_{-383}^{+401}$ & 1.358 & 23.9 & 0.302 & 0.155 & 0.065 & 386 & 5.23 & 0.9/4.51/9.02 & 11.56/57.81/115.63 \\
2M08383283+1946256 & K2-104 b & 641$_{-233}^{+273}$ & 0.875 & 9.3 & 0.391 & 0.204 & 0.024 & 736 & 69.53 & 4.45/22.25/44.51 & 57.03/285.14/570.28 \\
2M09052674+2140075 & K2-344 b & 343$_{-106}^{+112}$ & 0.566 & … & 0.475 & 0.25 & … & … & … & …/…/… & …/…/… \\
2M09533093+3534171 & Wolf 327 b & 331$_{-151}^{+201}$ & 0.316 & 44.4 & 0.301 & 0.155 & 0.01 & 986 & 224.08 & 8.94/44.71/89.41 & 114.56/572.81/1145.62 \\
2M10302934+0651492 & K2-324 b & 674$_{-187}^{+195}$ & 0.973 & … & 0.408 & 0.212 & 0.033 & 635 & 38.46 & 2.52/12.58/25.16 & 32.23/161.17/322.34 \\
2M10374104+0617094 & K2-323 b & 188$_{-84}^{+107}$ & 0.299 & … & 0.429 & 0.223 & 0.128 & 332 & 2.87 & 0.05/0.26/0.52 & 0.67/3.34/6.68 \\
2M18543080+4823277 & Kepler-1651 b & 518$_{-173}^{+192}$ & 0.748 & 18.18 & 0.389 & 0.202 & 0.062 & 452 & 9.9 & 0.55/2.77/5.53 & 7.09/35.43/70.86 \\
2M18545568+4557315 & Kepler-732 c & 192$_{-81}^{+100}$ & 0.268 & 36.14 & 0.4 & 0.208 & 0.014 & 966 & 206.49 & 3.88/19.39/38.78 & 49.68/248.42/496.83 \\
2M18545568+4557315 & Kepler-732 b & 192$_{-81}^{+100}$ & 0.268 & 36.14 & 0.4 & 0.208 & 0.068 & 439 & 8.83 & 0.17/0.83/1.66 & 2.12/10.62/21.25 \\
2M18545777+4730586 & Kepler-617 b & 718$_{-218}^{+232}$ & 1.193 & 38.27 & 0.441 & 0.23 & 0.022 & 815 & 104.19 & 7.11/35.57/71.14 & 91.15/455.73/911.47 \\
2M18575437+4615092 & Kepler-1074 b & 532$_{-142}^{+153}$ & 1.019 & 30.65 & 0.522 & 0.275 & 0.053 & 572 & 25.4 & 1.03/5.14/10.28 & 13.17/65.83/131.67 \\
2M18594123+4558206 & Kepler-504 b & 759$_{-255}^{+284}$ & 0.66 & 44.16 & 0.303 & 0.157 & 0.064 & 392 & 5.6 & 0.45/2.27/4.55 & 5.83/29.14/58.28 \\
2M19000314+4013147 & Kepler-974 b & 806$_{-153}^{+155}$ & 1.278 & 35.25 & 0.423 & 0.22 & 0.04 & 592 & 29.05 & 2.32/11.61/23.21 & 29.74/148.69/297.39 \\
2M19023192+7525070 & TOI-2095 c & 576$_{-176}^{+185}$ & 0.693 & 40 & 0.369 & 0.192 & 0.138 & 296 & 1.82 & 0.1/0.52/1.04 & 1.33/6.65/13.31 \\
2M19023192+7525070 & TOI-2095 b & 576$_{-176}^{+185}$ & 0.693 & 40 & 0.369 & 0.192 & 0.101 & 346 & 3.38 & 0.19/0.96/1.93 & 2.47/12.33/24.67 \\
2M19034293+3831155 & Kepler-1308 b & 324$_{-130}^{+152}$ & 0.423 & 33.95 & 0.365 & 0.189 & 0.023 & 722 & 64.49 & 2.33/11.63/23.26 & 29.8/149.01/298.03 \\
2M19062262+3753285 & Kepler-1124 b & 535$_{-170}^{+186}$ & 0.729 & 28.73 & 0.414 & 0.217 & 0.031 & 664 & 46 & 2.16/10.82/21.64 & 27.73/138.66/277.32 \\
2M19092321+4746226 & Kepler-1049 b & 611$_{-127}^{+132}$ & 1.163 & 29.17 & 0.52 & 0.275 & 0.033 & 720 & 63.63 & 2.95/14.77/29.53 & 37.84/189.21/378.41 \\
2M19130013+4640465 & Kepler-1350 c & 488$_{-183}^{+207}$ & 0.68 & 19.3 & 0.415 & 0.217 & 0.022 & 784 & 89.6 & 3.91/19.54/39.08 & 50.07/250.36/500.72 \\
2M19130013+4640465 & Kepler-1350 b & 488$_{-183}^{+207}$ & 0.68 & 19.3 & 0.415 & 0.217 & 0.041 & 575 & 25.89 & 1.13/5.65/11.29 & 14.47/72.34/144.68 \\
2M19213157+4317347 & Kepler-138 e & 444$_{-104}^{+114}$ & 0.698 & 18.86 & 0.465 & 0.245 & 0.18 & 293 & 1.73 & 0.06/0.3/0.61 & 0.78/3.9/7.8 \\
2M19213157+4317347 & Kepler-138 d & 444$_{-104}^{+114}$ & 0.698 & 18.86 & 0.465 & 0.245 & 0.128 & 347 & 3.45 & 0.12/0.61/1.21 & 1.55/7.77/15.53 \\
2M19213157+4317347 & Kepler-138 c & 444$_{-104}^{+114}$ & 0.698 & 18.86 & 0.465 & 0.245 & 0.091 & 413 & 6.87 & 0.24/1.21/2.41 & 3.09/15.46/30.92 \\
2M19213157+4317347 & Kepler-138 b & 444$_{-104}^{+114}$ & 0.698 & 18.86 & 0.465 & 0.245 & 0.075 & 454 & 10.07 & 0.35/1.77/3.54 & 4.53/22.66/45.31 \\
2M19301848+3907151 & Kepler-1741 b & 615$_{-135}^{+139}$ & 0.851 & 29.79 & 0.433 & 0.228 & 0.03 & 691 & 53.86 & 2.66/13.31/26.62 & 34.1/170.52/341.04 \\
2M19312949+4103513 & Kepler-45 b & 580$_{-147}^{+154}$ & 1.209 & 15.8 & 0.535 & 0.281 & 0.03 & 775 & 85.29 & 3.93/19.64/39.27 & 50.32/251.61/503.22 \\
2M19543665+4357180 & Kepler-186 f & 358$_{-141}^{+166}$ & 0.551 & 33.69 & 0.452 & 0.238 & 0.393 & 195 & 0.34 & 0.01/0.05/0.1 & 0.13/0.65/1.29 \\
2M19543665+4357180 & Kepler-186 e & 358$_{-141}^{+166}$ & 0.551 & 33.69 & 0.452 & 0.238 & 0.122 & 350 & 3.57 & 0.11/0.53/1.05 & 1.35/6.74/13.47 \\
2M19543665+4357180 & Kepler-186 d & 358$_{-141}^{+166}$ & 0.551 & 33.69 & 0.452 & 0.238 & 0.086 & 417 & 7.13 & 0.21/1.05/2.1 & 2.69/13.45/26.9 \\
2M19543665+4357180 & Kepler-186 c & 358$_{-141}^{+166}$ & 0.551 & 33.69 & 0.452 & 0.238 & 0.058 & 510 & 16.02 & 0.47/2.36/4.72 & 6.05/30.23/60.46 \\
2M19543665+4357180 & Kepler-186 b & 358$_{-141}^{+166}$ & 0.551 & 33.69 & 0.452 & 0.238 & 0.038 & 628 & 36.88 & 1.09/5.43/10.86 & 13.92/69.58/139.16 \\
2M20004946+4501053 & Kepler-560 b & 708$_{-318}^{+419}$ & 0.576 & 50.47 & 0.291 & 0.151 & 0.092 & 321 & 2.52 & 0.19/0.97/1.94 & 2.49/12.46/24.92 \\
\enddata
\tablenotetext{a}{IHZ derived for planetary masses of 1 M$_{\oplus}$.}
\tablenotetext{b}{For r$_{\rm M}$=5R$_{p}$, considering f=1$\%$, 5$\%$ and 10$\%$.}
\tablenotetext{c}{For r$_{\rm M}$=11.7R$_{p}$, considering f=1$\%$, 5$\%$ and 10$\%$.}
\end{deluxetable*}

\acknowledgments

K.C. thanks Jeremy Drake for helpful discussions. F.W. acknowledges support from fellowship by Coordena\c c\~ao de Ensino Superior - CAPES. 
K.C. and V.S. acknowledge that their work here is supported,
in part, by the National Science Foundation through NSF grant No. AST-2009507.
O.K. acknowledges support by the Swedish Research Council (grant agreements no. 2019-03548 and 2023-03667), the Swedish National Space Agency, and the Royal Swedish Academy of Sciences.
D.S. thanks the National Council for Scientific and Technological Development – CNPq.
CAP acknowledges financial support from the Spanish Ministry MICINN projects AYA2017-86389-
P and PID2020-117493GB-I00. SM Acknowledges funding from NASA XRP Grant 80NSSC24K0155. The Center for Exoplanets and Habitable Worlds is supported by the Pennsylvania State University and the Eberly College of Science.
Funding for the Sloan Digital Sky Survey IV has been provided by the Alfred P. Sloan Foundation, the U.S. Department of Energy Office of Science, and the Participating Institutions. SDSS-IV acknowledges support and resources from the Center for High-Performance Computing at the University of Utah. The SDSS web site is www.sdss.org.
SDSS-IV is managed by the Astrophysical Research consortium for the Participating Institutions of the SDSS Collaboration including the Brazilian Participation Group, the Carnegie Institution for Science, Carnegie Mellon University, the Chilean Participation Group, the French Participation Group, Harvard-Smithsonian Center for Astrophysics, Instituto de Astrof\'isica de Canarias, The Johns Hopkins University, 
Kavli Institute for the Physics and Mathematics of the Universe (IPMU) /  University of Tokyo, Lawrence Berkeley National Laboratory, Leibniz Institut f\"ur Astrophysik Potsdam (AIP),  Max-Planck-Institut f\"ur Astronomie (MPIA Heidelberg), Max-Planck-Institut f\"ur Astrophysik (MPA Garching), Max-Planck-Institut f\"ur Extraterrestrische Physik (MPE), National Astronomical Observatory of China, New Mexico State University, New York University, University of Notre Dame, Observat\'orio Nacional / MCTI, The Ohio State University, Pennsylvania State University, Shanghai Astronomical Observatory, United Kingdom Participation Group,
Universidad Nacional Aut\'onoma de M\'exico, University of Arizona, University of Colorado Boulder, University of Oxford, University of Portsmouth, University of Utah, University of Virginia, University of Washington, University of Wisconsin, Vanderbilt University, and Yale University.

\facility {Sloan}

\software{Matplotlib (\citealt{Hunter2007_matplotlib}), Numpy (\citealt{harris2020_numpy}), Synmast (\citealt{kochukhov2010_synmast}), Turbospectrum (\citealt{plez2012_turbospectrum})}

\bibliographystyle{yahapj}
\bibliography{references.bib}

\end{document}